\documentclass[preprint,aps,showpacs,amsfonts,epsf]{revtex4}

\input{epsf.tex}

\newcommand{\be}{\begin{equation}}
\newcommand{\ee}{\end{equation}}
\newcommand{\bea}{\begin{eqnarray}}
\newcommand{\eea}{\end{eqnarray}}
\newcommand{\ba}{\begin{array}}
\newcommand{\ea}{\end{array}}
\def\J#1#2#3#4{{#1} {\bf #2}, #3 (#4)}
\def\PRD{Phys. Rev. D}
\def\PR{Phys. Rev.}
\def\PRL{Phys. Rev. Lett.}

\def\JMP{J. Math. Phys.}

\begin{document}
\draft
\title{Comment on\\``Global Structure of the Kerr Family of Gravitational Fields''}
\author{H. Garc\'ia-Compe\'an and V.~S.~Manko}
\address{Departamento de F\'\i sica, Centro de Investigaci\'on y de
Estudios Avanzados del IPN, A.P. 14-740, 07000 M\'exico D.F.,
Mexico}

\begin{abstract}
We comment on the role of the Cartesian-type Kerr-Schild
coordinates in developing a faulty maximal extension of the
Kerr-Newman solution in the well-known paper of Carter.
\end{abstract}

\pacs{04.20.Jb, 04.70.Bw, 97.60.Lf}

\maketitle


In his famous paper \cite{Car}, Carter generalized the results of
Boyer and Lindquist \cite{BLi}, obtained for the Kerr solution
\cite{Ker}, to the case of the Kerr-Newman (KN) spacetime
\cite{NCC}. Among various important discoveries made in
\cite{Car}, one should of course mention the writing of the KN
metric in the canonical form with only one non-diagonal component,
and a complete integration of the Hamilton-Jacobi equation for the
geodesics. At the same time, Carter's approach to the maximal
extension of the KN solution, which involves, like in the Kerr
case \cite{BLi}, a continuation of the radial coordinate $r$ to
infinite negative values and the essential use of the disk
geometry of the surface $r=0$, $t={\rm constant}$ (henceforth
referred to as the $r=0$ surface for simplicity) has been recently
invalidated by the rigorous proof \cite{GMa} that the latter
two-surface is not a disk. The present comment deals with
answering an interesting and natural question of why for nearly
fifty years the surface $r=0$ of the KN solution has been given a
wrong interpretation.

Suppose we have some stationary solution of the field equations
defined by a metric written in the Boyer-Lindquist-like
coordinates $\{r,\theta,\varphi,t\}$, whose surface $r=0$ may in
principle have an arbitrary geometry yet to be identified. Then,
with the idea to ``attain a better insight'' of the geometrical
properties of the solution, let us make a coordinate change,
introducing the Cartesian-type coordinates $\{x,y,z,t'\}$ via the
formulas
\be x+iy=(r+ia)e^{i\varphi'(\varphi,r)}\sin{\theta}, \quad
z=r\cos{\theta}, \quad t'=t'(t,r), \label{trans} \ee
where $a$ is a rotation parameter. An immediate corrolary of
(\ref{trans}) are two equations,
\be x^2+y^2=(r^2+a^2)\sin^2\theta  \quad\hbox{and} \quad
z^2=r^2\cos^2\theta, \label{norm} \ee
whence we first readily obtain the relation
\be \frac{x^2+y^2}{r^2+a^2}+\frac{z^2}{r^2}=1, \label{ellips} \ee
which, when $r\ne0$, shows that the surfaces $r={\rm constant}$
are confocal ellipsoids, and then, in the limit $r=0$, we yield
the equations
\be x^2+y^2=a^2\sin^2\theta, \quad z=0, \label{disk} \ee
defining a disk of radius $|a|$ in the equatorial $z=0$ plane.

Therefore, the coordinate transformation (\ref{trans}), which was
first proposed in Kerr's celebrated paper \cite{Ker} and nowadays
better known under the name of the Kerr-Schild transformation, is
a remarkable invention indeed, as it permits one to convert in a
magic way any unknown surface $r=0$ of arbitrary geometry into a
disk (\ref{disk}), without even writing a concrete metric
explicitly! It is no surprise, then, that the above very simple
and seemingly efficient procedure of identifying the surface $r=0$
as a disk was employed not only for the analysis of the Kerr
geometry by Boyer and Lindquist \cite{BLi}, but also by Carter in
the KN asymptotically flat case \cite{Car}, and in the
asymptotically non-flat case of stationary black holes in the
presence of the cosmological constant (see, e.g., \cite{Car2}). We
would like to emphasize that, surprisingly, nobody has put in
doubt the absolutely formal ``disk'' interpretation of the surface
$r=0$ springing up from the transformation (\ref{trans}).

It is remarkable that in the case of the KN solution the disk
geometry of the two-surface $r=0$ can be easily refuted by
calculating its Gaussian curvature $K$; this has been first done
in our paper \cite{GMa}, resulting in the following expression:
\be K=Q^2F(\theta), \label{K} \ee
where $Q$ is the charge parameter, and $F(\theta)$ is a function
of $\theta$ whose explicit form may be found in \cite{GMa}. Since
it is well known that the Gaussian curvature of a disk is equal to
zero, it is clear from (\ref{K}) that the Cartesian Kerr-Schild
coordinates actually fail to supply us with a correct description
of the surface $r=0$. It is worth mentioning that essentially all
stationary black-hole solutions with a cosmological constant turn
out to have the $r=0$ surface of nonzero Gaussian curvature
\cite{MGa}, thus providing additional examples of particular
nondisk geometries of that two-surface in the spacetimes different
from the KN one.

One might think, however, that at least in the case of the Kerr
solution the disk interpretation of the surface $r=0$ supplied by
the Cartesian coordinates must be correct because the Gaussian
curvature of the latter surface for the Kerr solution is known to
be equal to zero \cite{ONe} (this result also follows from
(\ref{K}) by setting $Q=0$). Unfortunately, even this ``very
obvious'' case is nothing more but a subtle mathematical puzzle,
the resolution of which requires the analysis of the surface $r=0$
in the Weyl-Papapetrou cylindrical coordinates and recalling that
a disk is not the only two-surface possessing zero Gaussian
curvature; then we must finally interpret the surface $r=0$ of the
Kerr solution as a dicone \cite{GMa}.

Therefore, the main conclusion that could be drawn from the above
consideration is that the Kerr-Schild coordinates (\ref{trans})
seem to bear principle responsibility for a fifty-year delay in
our understanding of the true global structure of the stationary
black-hole solutions. With the discovery of the genuine geometry
of the surface $r=0$ of the latter solutions, the elaboration of
the corresponding correct maximal analytic extensions will be,
hopefully, not long in coming.

\section*{Acknowledgments}

This work was partially supported by the Research Project~128761
from CONACyT of Mexico.

\end{document}